\documentclass[prb,twocolumn,superscriptaddress]{revtex4}

\usepackage{amsmath} 
\usepackage{amssymb} 
 \usepackage{amsfonts}

\usepackage{graphicx} 
\usepackage{footnote}

\usepackage{array}
\usepackage{multirow}

\begin{document}

\title{ Thermodynamic Description of the LCST of Charged Thermoresponsive Copolymers} 

\author{Jan Heyda}
\affiliation{Soft Matter and Functional Materials, Helmholtz-Zentrum Berlin, Hahn-Meitner Platz 1, 14109 Berlin, Germany}

\author{Sebastian Soll}
\affiliation{Department of Colloid Chemistry, Max-Planck-Institute of Colloids and Interfaces, D-14476 Potsdam, Germany}

\author{Jiayin Yuan}
\affiliation{Department of Colloid Chemistry, Max-Planck-Institute of Colloids and Interfaces, D-14476 Potsdam, Germany}
\email{Jiayin.Yuan@mpikg.mpg.de}

\author{Joachim Dzubiella}
\email{joachim.dzubiella@helmholtz-berlin.de}
\affiliation{Department of Physics, Humboldt-University Berlin, Newtonstr.~15, 12489 Berlin, Germany}
\affiliation{Soft Matter and Functional Materials, Helmholtz-Zentrum Berlin, Hahn-Meitner Platz 1, 14109 Berlin, Germany}

\begin{abstract}
The dependence of the lower critical solution temperature (LCST) of charged, thermosensitive copolymers on their charge fraction and the salt concentration is investigated by employing systematic cloud-point experiments and analytical theory. The latter is based on the concept of the Donnan equilibrium incorporated into a thermodynamic expansion of a two-state free energy around a charge-neutral reference homopolymer and should be applicable for weakly charged (or highly salted) polymer systems.  Very good agreement is found between the theoretical description and the experiments for aqueous solutions of the responsive copolymer poly(NIPAM\textit{-co-}EVImBr) for a wide range of salt concentrations and charge fractions up to 8\%, using only two global, physical fitting parameters.  
\end{abstract}

\maketitle

In the last years we have witnessed a massive increase in the development of stimuli-responsive copolymers with 
various architectures for the potential use in 'smart' functional materials, such as drug carriers, antifouling coatings, soft biomimetical tissue, 
nanoreactors, and carbon nanotube dispersions.~\cite{stuart,bajpai,liechty,gil,qiu,alarcon,Jiayin,wu,NewThermoPolym} 
The virtue of stimuli-responsive copolymers is that close to their lower critical solution temperature (LCST), a stimulus can 
easily switch the copolymer's hydration properties from being mainly hydrophilic to hydrophobic leading to substantial changes
in its physicochemical properties.  Thus, near the LCST the material is easily tuneable to 
possess the functional behavior adequate for a desired application.   

A very important tool to control the LCST of a polymer is the copolymerization to introduce electrostatically charged groups and the subsequent fine-tuning by salt concentration and the solution's pH value.~\cite{khoklov, suzuki, yoo, jones,Kawasaki,KawasakiJPCB1997,KawasakiJPCB1997A,Principi2000,Spafford1998,Karbarz2006,Szczubialka,Mori2004} Analytical formulas which describe and predict LCST changes with charge fraction and salt concentration could be highly useful for the guidance of these effects. In this contribution we present a starting point for such a theoretical  framework for slightly charged polymers and compare to  novel, well-controlled and systematic LCST estimates by cloud-point measurements of the responsive copolymer poly(N-isopropylacrylamide\textit{-co-}1-ethyl-3-vinylimidazolium bromide); poly(NIPAM\textit{-co-}EVImBr).\cite{Jiayin}  We find excellent agreement of our theoretical approach with the experiments for a large parameter space covering charge fractions up to 8\% and the millimolar to molar range of salt concentrations. Our description may serve as a guide for the optimization of stimuli-responsive copolymer architectures in the future design of soft functional materials. 

In our  theoretical model we assume that 
the copolymer volume transition at the LCST can  be understood as a transition from a dense collapsed state of the copolymer to an expanded, coil-like 
state in a bimodal free energy landscape~\cite{kawaguchi} as a function of polymer size, for instance, the copolymer specific volume $v$. 
Schematically, the free energy $G(v)$ of such a two-state model is  shown in Fig.~1, with the corresponding minima in the compact globular (g) and the extended coil (c) 
states at $v_g$ and $v_c$, respectively. 
\begin{figure}[h!]
\begin{center}
\includegraphics[width=8.0cm,angle=0]{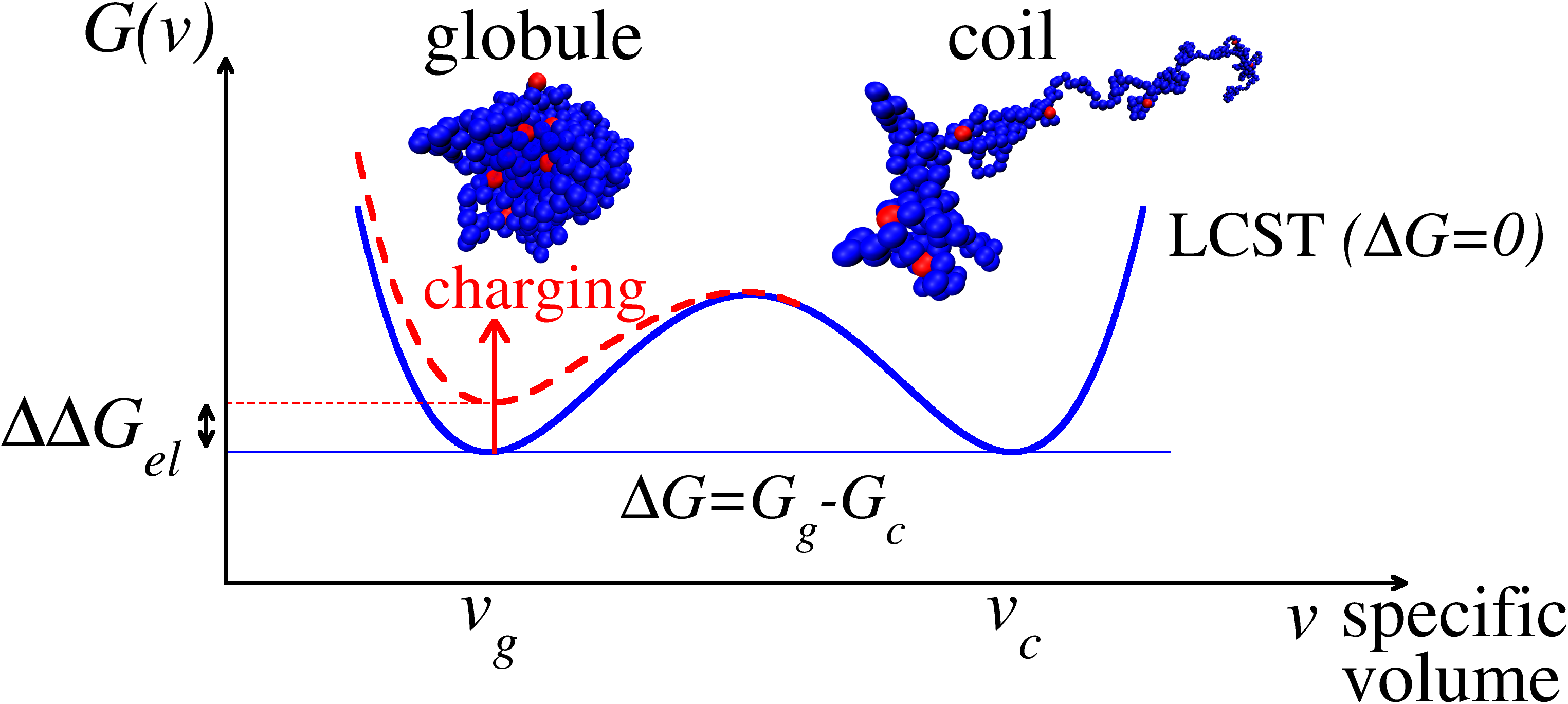}
\caption{Schematic free energy landscape $G(v)$ of a copolymer in a two-state model as function of the specific volume $v$. At the LCST the collapsed (globule) and extended (coil) states at $v_g$ and $v_c$, respectively, of the neutral copolymer are equally probable and $\Delta G=G_g-G_c=0$ (solid blue curve). The free energy of the compact globule state is elevated by a small amount $\Delta\Delta G_{el}(\alpha; c_s)$ by charging a small fraction $\alpha$ of copolymer monomers (dashed red curve), thus favoring the coil over the globule state. This perturbation can be compensated by a change $\Delta T$ of the temperature leading to a new LCST. In the illustrative conformations, the blue balls denote neutral polymer beads, while red balls are charged beads.}
\label{Schematic_figure_DG_landscape}
\end{center}
\end{figure}

Right at the LCST of the neutral homopolymer, in the following denoted as $T_0$, the free energy difference (per monomer) $\Delta G(T_0)$
between collapsed and coil states vanishes, that is, $\Delta G(T_0) = G_g(T_0)-G_c(T_0)=0$. For the latter equation to hold we assume very narrow free energy minima such that in equilibrium $G(v)$ is dominated by its values right at $v_c$ and $v_g$, respectively.  The transition is accompanied by a transition entropy  per monomer 
$\Delta S_0\equiv \Delta S(T_0)$. If we identify $G$ with the Gibbs free energy, then the usual thermodynamic relation 
$\Delta G = \Delta H - T \Delta S$ provides $\Delta S_0 = \Delta H(T_0)/T_0$ at the LCST, where
$\Delta H(T_0)$ is the corresponding transition enthalpy per monomer. 

Charging the copolymer by a small fraction $\alpha=|z\chi|$, defined by the charge valency $z$ times $\chi$, the ratio  between the numbers of chargeable copolymer monomers and total monomers $N$, 
will lead to a perturbation of the bimodal free energy distribution.  In the following we denote $\alpha$ 
as {\it charge fraction} and $\chi$ as the {\it copolymer fraction}.  For weak charging the
effects in the compact globule will be much larger than for the expanded state due to the close spatial proximity of the
charged groups and the resulting electrostatic repulsion.  Let us assume in  the following that charging elevates the compact state by an electrostatic 
energy $\Delta\Delta G_{el}(\alpha; c_s)$, while the expanded state remains unperturbed, cf. the illustration in Fig.~1. 
This small shift $\Delta\Delta G_{el}(\alpha; c_s)$ will be a function of charge fraction $\alpha$ and the salt concentration 
$c_s$ and can be  reversed by a small temperature change $\Delta T$, which controls the equilibrium weight between 
coil and globule states.  For small temperature changes and weak copolymerization,  $\Delta G(T,\chi,\alpha,c_s)$ 
can now be expressed by a Taylor-expansion viz.
\begin{eqnarray}
\nonumber
\Delta G(\Delta T,\chi,\alpha,c_s)\!\! &\simeq& \!\! \Delta G(T_0) +\!\! \left.\frac{\partial \Delta G}{\partial T}\right\vert_{0}\!\!\Delta T +\!\! \left.\frac{\partial^2 \Delta G}{2\partial T^2}\right|_{0}\!\!\Delta T^2 \\ \nonumber 
&+& \left.\frac{\partial \Delta G}{\partial \chi}\right\vert_{0}\chi+\left.\frac{\partial^2 \Delta G}{\partial T\partial \chi}\right|_{0}\Delta T\chi \\
 &+&\left.\frac{\partial \Delta G}{\partial c_s}\right\vert_{0}c_s +\Delta\Delta G_{el}(\alpha; c_s),
\end{eqnarray}
where the subscript '0' denotes the expansion around the electroneutral and salt-free reference state of the homopolymer at 
$T_0, \chi=0$, $c_s=0$.  In our study this reference state is the pure poly(N-isopropylacrylamide) (PNIPAM) homopolymer. 
We expanded to second order in $T$. As will become clearer in the remainder of the 
manuscript, it is also important to expand $\Delta G$ in specific copolymerization $\chi$ and salt concentration $c_s$. The expansion in $\chi$ and $c_s$ considers {\it nonelectrostatic} effects on $\Delta G$ due to the specific change in chemical solvation thermodynamics upon copolymerization or the addition of salt, respectively. In other words, 
even for a neutral copolymer, copolymerization and the addition of salt will specifically change the LCST. The reasoning behind our expansion is illustrated also in the SI for the convenience of the reader. All nonspecific, {\it purely electrostatic} charging and salt screening effects are included {\it nonlinearly} in the term $\Delta\Delta G_{el}(\alpha; c_s)$ which will be defined later.    

Using the thermodynamic relations $\Delta S_0 = - {\partial \Delta G(T_0)} /{\partial T}$ and ${\Delta C_p} = - T_0 {\partial^2 \Delta G(T_0)}/{\partial T^2}$, 
where ${\Delta C_p}$ is the isobaric transition heat capacity, and defining the thermodynamic slopes ${\partial \Delta G}/{\partial \chi}=m_\chi$, ${\partial \Delta G}/{\partial c_s}=m$, and
${\partial \Delta S_0}/{\partial \chi}=m'_\chi$, and enforcing  $\Delta G(\Delta T,\chi,\alpha,c_s)=0$ at the new LCST with value $T_0+\Delta T$, it is
\begin{eqnarray}
(\Delta S_0+m'_\chi\chi)\Delta T   &+&   \frac{{\Delta C_p}}{2T_0}\Delta T^2 \\
&\simeq& \Delta\Delta G_{el}(\alpha;c_s) +  m_\chi \chi +  m c_s. \nonumber
\end{eqnarray}
We rearrange  to obtain an explicit expression for the change in LCST  
\begin{eqnarray}
\Delta T \simeq  \frac{\Delta\Delta G_{el}(\alpha;c_s)+m_\chi\chi+mc_s}{\Delta S_0+m'_\chi\chi}, 
\end{eqnarray}
where we have neglected heat capacity effects.  The latter is a valid measure for small changes 
$\Delta T\ll T_0$ as is the case in our study. The full equation including a discussion of 
heat capacity effects is shown and discussed in the supplementary information (SI).

We now estimate the purely electrostatic free energy change $\Delta\Delta G_{el}(\alpha; c_s)$ upon charging the copolymer in the solution with monovalent salt concentration $c_s$. We make 
the mean-field assumption that the  collapsed copolymer globules can be modelled as homogeneous, compact entities with a mean  monomer number density $\rho_g=N/V_g$ and the salty environment will readily lead to a complete electrostatic neutralization of the globule. Hence, the free energy change is just based on the changes of ionic osmotic pressure, 
called the Donnan pressure.  Given the globule with volume $V_g$  in contact with the salty reservoir at constant chemical potential, 
the free energy change per monomer is given by  
\begin{eqnarray}
 \Delta\Delta G_{el} (\alpha;c_s)=  \Delta P(\alpha;c_s) V_g/N = \Delta P(\alpha;c_s)/ \rho_g,  
 \end{eqnarray}
where $k_BT = \beta^{-1}$ is the thermal energy,  and $\Delta P(\alpha;c_s) = P_g(\alpha;c_s)-P_b$ is the net 
osmotic pressure inside the globule, that is, the difference between osmotic pressure inside
the globule and bulk, respectively. For a simple salt at low concentration 
the ions can be well treated as an ideal gas and thus  $\beta \Delta P(\alpha;c_s) = c_+ + c_--2c_s$, 
where 
\begin{eqnarray}
c_+  = c_s\exp(-e\beta\Delta\phi) \;\;{\rm and}\;\; c_-  = c_s\exp(e\beta\Delta\phi)
\end{eqnarray}
are the cation and anion concentration inside the globule, respectively, determined by the Donnan 
potential $\Delta\phi$. The latter describes the difference between the electrostatic potentials in 
the globule and the reference bulk solution. It is determined by the electroneutrality condition $c_+-c_-+\alpha\rho_g=0$
which leads to 
\begin{eqnarray}
\beta\Delta\phi  = \ln (y+\sqrt{y^2+1}), 
\end{eqnarray}
where $y= \alpha  \rho_g/(2 c_s)$. With that we finally find for the purely electrostatic 
free energy change
\begin{eqnarray}
\Delta\Delta G_{el}(\alpha;c_s) = \frac{2k_BT_0c_s}{\rho_g}\left({\sqrt{y^2+1}-1}\right). 
\end{eqnarray}
The  final result 
for the change $\Delta T$ of the LCST in a monovalent salt reads
\begin{eqnarray}
\Delta T \simeq  \frac{2k_BT_0c_s}{\Delta S(\chi)\rho_g}\!\left(\!{\sqrt{y^2+1}-1}\right)\!+\!\frac{m_\chi\chi}{\Delta S(\chi)}+\frac{mc_s}{\Delta S(\chi)}, 
\label{1st_order}
\end{eqnarray} 
where we abbreviated $\Delta S(\chi)=\Delta S_0+m'_\chi\chi$.
Eq.~(\ref{1st_order}) is the key equation derived in this work. For high salt  concentrations or very small charge fractions ($y\ll1)$, eq.~(\ref{1st_order})
further simplifies to 
\begin{eqnarray}
\Delta T \simeq  \frac{k_BT_0  \rho_g\alpha^2}{4\Delta S(\chi) c_s} +\!\frac{m_\chi\chi}{\Delta S(\chi)}+\frac{mc_s}{\Delta S(\chi)}. 
\label{1st_order_ll}
\end{eqnarray} 
In the limit of very weak copolymerization and no ion-specific effects we can set $m_\chi=0$, $m'_\chi=0$, and $m=0$, 
and one obtains a quadratic scaling with charge fraction $\alpha$ and an inversely linear scaling 
in $c_s$ as  
\begin{eqnarray}
\Delta T \simeq  \frac{k_BT_0  \rho_g\alpha^2}{4\Delta S_0 c_s} 
\label{llw}
\end{eqnarray}
which represents a limiting law for very weak perturbations 
by charge effects only.  Interestingly, it has the same scalings as 
mean-field descriptions of the second virial coefficient of 
polyelectrolytes.~\cite{takahashi}

Apart from the electrostatic 'external' control parameters, such as the 
salt concentration $c_s$ and the charge fraction $\alpha$, the other 
determinants in the theory are intrinsic material constants specific to the 
{\it neutral} reference homopolymer, such as the  LCST $T_0$, 
the transition entropy per monomer  $\Delta S_0$, 
the globule density $\rho_g$, and the specific changes induced by 
copolymerization and ions as expressed by $m_\chi$, $m'_\chi$ and $m$, 
respectively. We discuss experimental values of those and our fitting 
procedure in the following. 

The nonelectrostatic ion-specific effects for {\it simple} and monovalent salts are typically linear in salt concentration,~\cite{pnipam:lcst,cremer:2007} that is $\propto mc_s$, as already included in our expansion in eq.~(1). Higher order ion-specific effects are not important in this work but could be for more complex salts.
The so-called $m$-value is the coefficient which describes the change of transition free energy per monomer with salt concentration and is a salt-specific number.~\cite{record} For our salt KBr and pure PNIPAM, which is our reference polymer, this $m$-value has been measured and is $m=-111$\,J/mol/(mol/l), i.e. to the slope $\Delta T / \Delta c_s =-7.4$ \,K/(mol/l) and $\Delta S_{0}=15\, {\rm J/mol/K}$.~\cite{pnipam:lcst,cremer:2007} For our poly(NIPAM\textit{-co-}EVImBr) we calculated 
this slope from the data at high salt concentrations ($c_s>0.2$~M), where nonspecific charge effects have vanished, 
and find $m=-105 \pm 15 $~\,J/mol/(mol/l) in very good agreement with the one in the 
pure PNIPAM system. The found $m$-value stays fixed in our fitting procedure. 

From subtracting the specific ion effects ($\propto mc_s$)  from our experimental data we find that the LCST  
is $T_0=306\pm1$~K for the PNIPAM homopolymer 
reference state, independent of the the particular choice of copolymer fraction 
$\chi$. This LCST value is very close to the experimentally known LCST of pure PNIPAM of about $T_0=305\pm1$~K.~\cite{gil} We conclude that  specific copolymerization effects on the free energy are small for our system, that is, $m_\chi\simeq 0$, and can be neglected. We keep the value of $T_0=306$~K fixed in our global fitting procedure.

The transition entropy per monomer $\Delta S_{0}$ for pure PNIPAM chains has been experimentally measured for various chain length 
with data summarized in  Tab.~I.  The spread in the data is relatively narrow and we fix $\Delta S_0$ in our fitting procedure 
to the mean experimental value $\Delta S_0=15$~J/mol/K.  Previous experiments, however, show a very sensitive dependence of $\Delta S_0$ on functionalization.~\cite{Winnik2013, KawasakiJPCB1997,Yamazaki2000} This change is expressed in the parameter $m'_\chi=\partial \Delta S_0/\partial \chi$ and depends on the specific chemical  nature of the copolymerization.  Since measurements are unavailable, we will use $m'_\chi$ as one of the global fitting parameters in our study. 

\begin{table}[ht]
\begin{center}
\begin{tabular}{ c | l }
$\Delta S_{0}$ &  \\
$\rm [J/mol/K] $& reference/comment \\
\hline 
17.0  & \cite{Kato} \\
 6.0  & \cite{Freitag} 20-mer of PNIPAM \\
 12.4  & \cite{Ptitsyn1995} 92-mer of PNIPAM \\
 18.4  & \cite{Ptitsyn1994} 600-mer of PNIPAM \\
 15.3   & \cite{Kawasaki} \\
 17.8  & \cite{Kujawa2005} C$_{18}$-PNIPAM-C$_{18}$ \\
 19.2  & \cite{Kujawa2001} \\
 17.7     & \cite{Akashi2002} Microcalorimetry \\
 21.3  & \cite{Winnik2013} End group effects
\end{tabular} 
\label{Table_label_SI1}
\caption{$\Delta S_{0}$ is the transition entropy per monomer of the pure PNIPAM homopolymer as collected from sources in the literature. Maximum errors are about $\pm$~2~J/mol/K.}
\end{center}
\end{table}

For pure PNIPAM the globular density $\rho_g$ was estimated previously to be in the molar range, roughly  
$3-6$\,M, based on the measurements of the hydrodynamic radius of the polymer in the collapsed state well 
above the LCST.~\cite{Wang1995, Wang1998}  We find, however, that with those values of $\rho_g$ the experimental data was difficult to fit in a comprehensive way. We suspect that this deficiency may result from our simplified two-state treatment of the copolymer free energy landscape. 
Experiments actually indicate that at coexistence the collapsed state may be only partially occupied by monomers due to the presence of 'crumpled coil'~\cite{Wu:prl} or perl-necklace structures.~\cite{pearl} Therefore, we use the {\it effective} globule number density $\rho_g$ as the 
second global fitting parameter.

Under the bottom line our approach has only two fitting parameters: the specific change of the transition entropy 
$\Delta S_0$ with copolymerization $\chi$ as expressed by the variable $m'_\chi$  and the effective monomer number 
density $\rho_g$ in the globular state which is expected to be in the molar range. With those constraints we perform 
a {\it global} fit,  i.e., the whole data set in our measured $\{\chi;c_s\}$ regime will be described by two single values 
of $m'_\chi$ and $\rho_g$,  respectively.  The reference LCST $T_0=306~$K, salt-specific corrections $m=105$~J/mol/(mol/l), 
and the transition entropy $\Delta S_0=15$~J/mol/K were kept fixed during the fitting. 

Our experiments provide the change in LCST by an amount $\Delta T(\chi,c_s)$ for a wide range of salt concentrations $c_s$ between 0.01 and 1 M and four copolymer fractions $\chi=1.2\%,3.1\%,4.8\%$, and 7.6\%. The latter correspond to the same absolute charge fractions $\alpha=|z\chi|$ since in our case $z=+1$. The changes in LCST are estimated by cloud point measurements (see Methods).  Since the volume transition from coil to aggregated globules  is not always sharp,~\cite{kawaguchi} the cloud point depends on the exact criterion how much light is transmitted through the solution.  Therefore we employ and compare the two criteria commonly used, i.e., the cloud point is defined by either 50\% or 90\% transmittance. 
Our model is expected to perform well for low charge fractions. Consistently we apply our global fitting only to data below a charge
fraction of 5\%. The fit for the 7.6\% is then a prediction of the theory.  The fitting  was performed by least 
square error minimization of the global fit by $m'_\chi$ and $\rho_g$ to the experimental data set 
$\{\alpha;c_s\}=\{0.012,0.031,0.048$; $0.015~{\rm M}, 0.020~{\rm M}, 0.040~{\rm M}, 0.1~{\rm M}\}$.

\begin{figure}[h!]
\begin{center}
\includegraphics[width=8.0cm,angle=0]{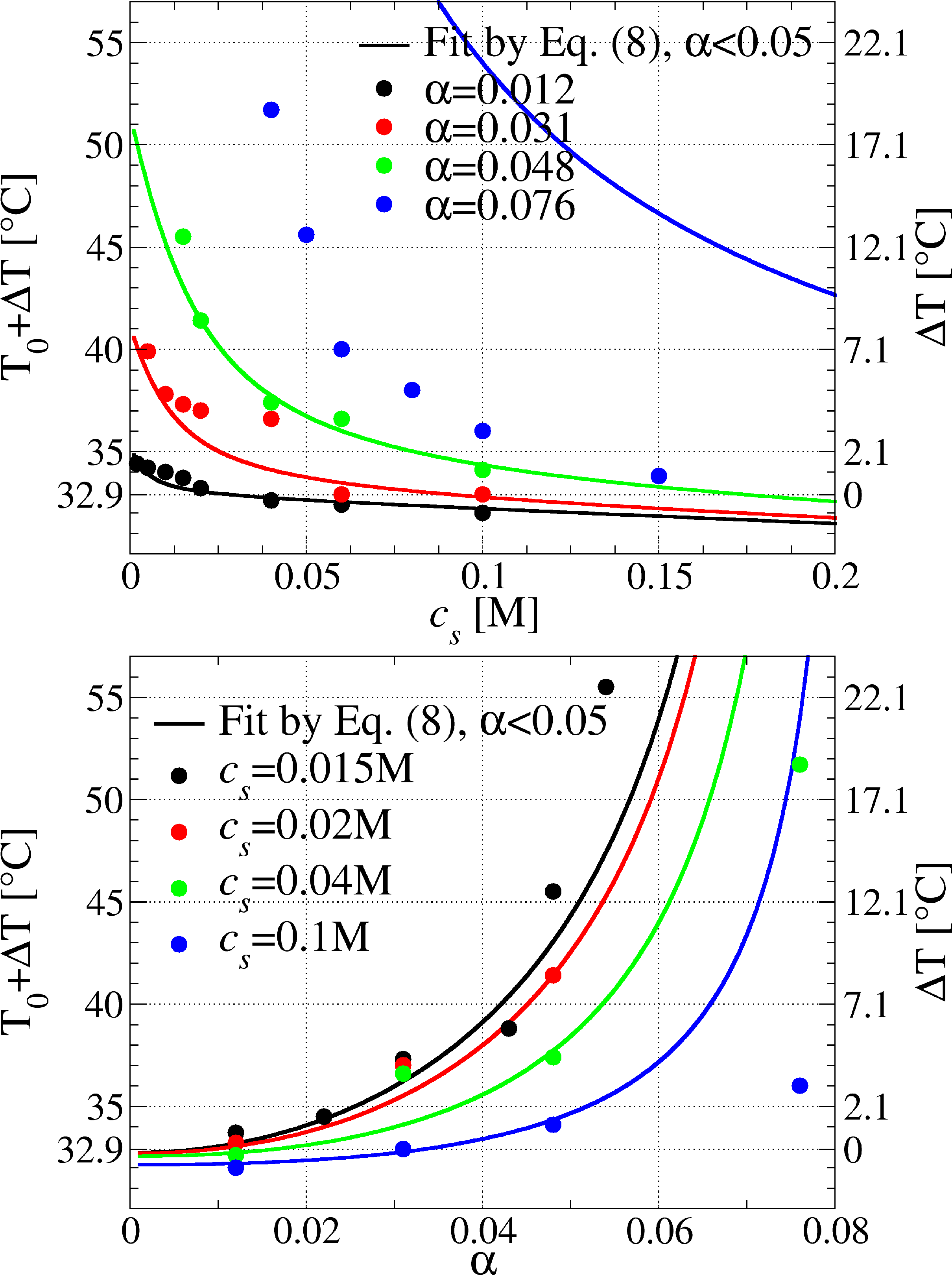}
\caption{LCST $T_0+\Delta T$ for all four copolymer fractions $\chi$ (i.e. charge fractions $\alpha$) as a function of salt concentration $c_s$  (upper panel), and for selected salt concentrations $c_s$  as a function of charge fraction $\alpha$  (lower panel). Filled circles denote experimental $T_0+\Delta T(\alpha;c_s)$ data derived with the 50\% transmittance criterion. Solid lines represent best theoretical fit based on eq.~(\ref{1st_order}) for $\alpha<5\%$. The fitting parameters are summarized in Tab.~II.}
\label{T_05_fit_figure}
\end{center}
\end{figure}

\begin{figure}[h!]
\begin{center}
\includegraphics[width=8.0cm,angle=0]{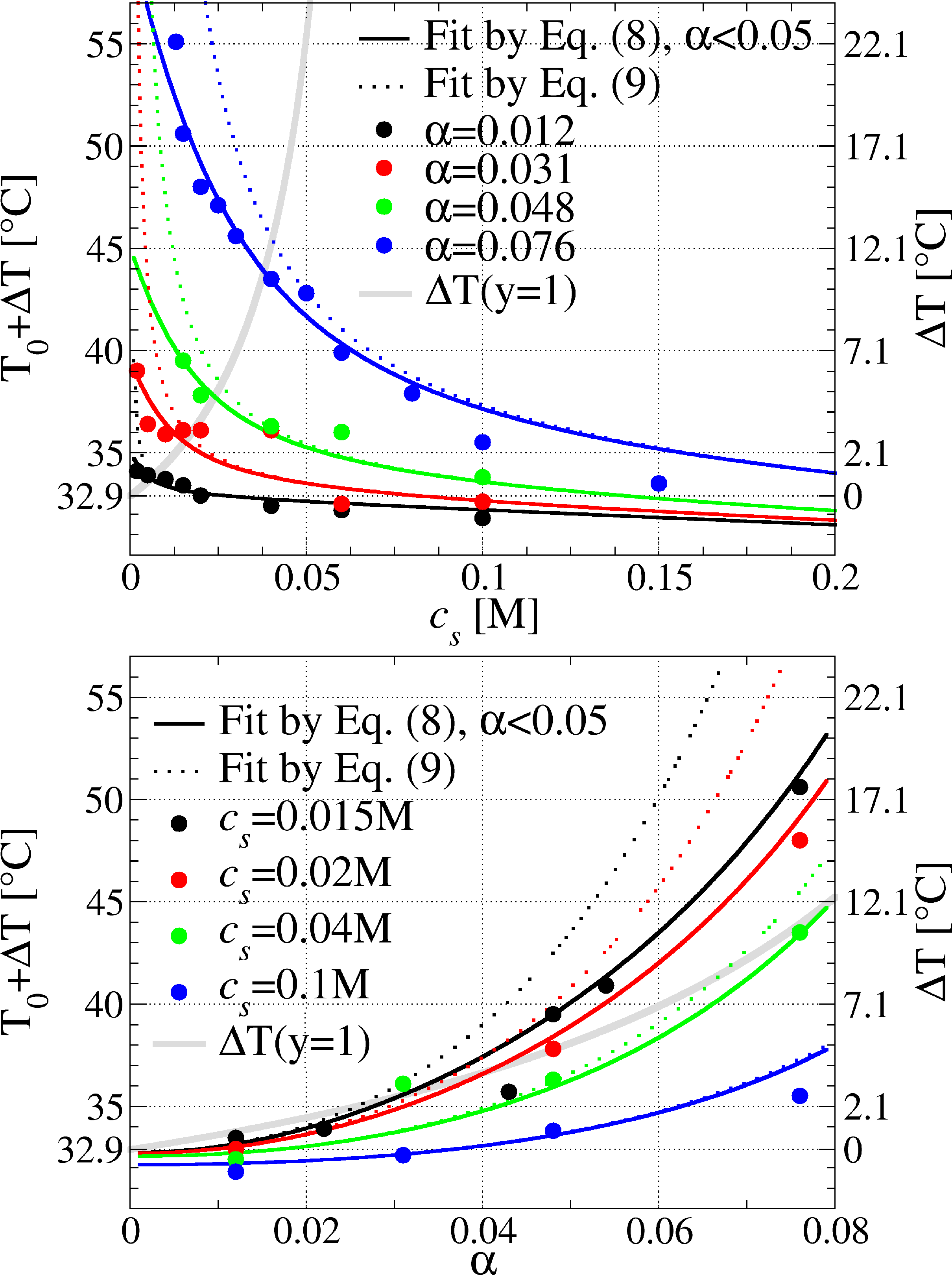}
\caption{LCST for selected copolymer fractions $\chi$ (i.e. charge fractions $\alpha$) as a function of salt concentration $c_s$  (upper panel), and for selected salt concentrations $c_s$ as a function of charge fraction $\alpha$ (lower panel) based on experimental data derived with the 90\% transmittance criterion. Solid lines represent best theoretical fit based on eq.~(\ref{1st_order}) for $\alpha<5\%$. The dotted lines are the corresponding curves for eq.~(\ref{1st_order_ll}) in the limit $y\ll1$. Fitting parameters are summarized in Tab.~II. The grey line represents the $\Delta T$-isocontour for $y=1$; according to eqs. (8, 9) in SI.}
\label{T_09_fit_figure}
\end{center}
\end{figure}

The experimentally measured $\Delta T (\alpha;c_s)$ data and the theoretical fits are presented in 
Figs.~\ref{T_05_fit_figure} and \ref{T_09_fit_figure} for the two cloud point criteria based on 
either 50\% or 90\%  transmittance, respectively.  The corresponding 
best-fit parameters $m'_\chi$ and $\rho_g$ are summarized in Tab.~\ref{Table_bestfit}. As we can 
see the fits work very well for the lower charge fractions $<5\%$ for both transmittance criteria. 
The data based on the 90\% transmittance criterion, however, 
is better described and even the highest charge fraction 7.6\% is well predicted, in contrast to the
50\% transmission data.  For the 90\% data the error of the fits is less than $\pm$ 1~K.  This may imply that estimates of the real 
LCST are better given by the 90\% transmission cloud point criterion than the 50\% one.
Overall the performance of the fitting for the 90\% transmission data is impressive given the fact that only two global
parameters (with physical meaning) have been employed. We have not found any 
improvement in the fitting by including heat capacity effects, see also the discussion in the SI. 
Note that the best fit of $\rho_g=1$~mol/l is smaller than experimental measurements of the globule density well above 
the LCST, ~\cite{Wang1995, Wang1998}  effectively reflecting conformations
due to a more complex free energy landscape than assumed in our model.

For the 90\% transmission data in Fig.~3 we have also plotted the simpler eq.~(\ref{1st_order_ll}) which should
be valid for $y\ll1$ (dotted lines).  As expected this form is only valid for the low charge 
fractions ($\alpha \lesssim 4.8\%$) or at high salt concentrations $c_s\gtrsim 0.04\,{\rm M}$. 
More generally, we find the equation describes well all data below $y=\alpha\rho_g/(2c_s)\simeq 1$, 
which is indicated by a thick grey curve in Fig.~3. The analytical expression for iso-curve 
$\Delta T(y=1)$ is provided in the SI. 

\begin{table}[h!]
\caption{The best-fit parameters obtained by eq.~(\ref{1st_order}) (lines in Figs. 2 and 3) and eq.~(\ref{llw})  (lines in Fig. 4) for the LCST determined for the T=50\% and at T=90\% transmittance criteria, respectively. The other relevant parameters are $\Delta S_0=15$\,J/mol/K, $m=-105$\,J/mol/(mol/l), and $T_0=306$\,K.}
\centering
\begin{footnotesize}
\begin{tabular}{ l | l | c | c }
Data & Theoretical fit & $m'_\chi$ $\left[ \frac{\rm J }{\rm mol \cdot K} \right]$ & $\rho_g$ $ \left[ \frac{\rm mol}{\rm l} \right] $ \\
\hline 
T=50\% & Eq.~(\ref{1st_order}), $\alpha \leq 4.8\%$, Fig.~2 &  -176 & 1.0 \\
T=90\% & Eqs.~(\ref{1st_order}),(\ref{1st_order_ll}), $\alpha \leq 4.8\%$, Fig.~3 &  -104  & 1.0 \\
T=90\% & Eq.~(\ref{llw}), all $\alpha$, Fig.~4                        	 	&  N.D. & 1.24 \\
\end{tabular}
\end{footnotesize}
\label{Table_bestfit}
\end{table}

Finally we demonstrate that for a first-and-quick pragmatic fit of the total data even the  simple limiting law eq.~(\ref{llw}) can be useful to describe the non-specific charging effects on the LCST. For this, the ion-specific slope $\propto m c_s$ is subtracted from the 90\% transmission data in Fig.~3 and presented in Fig.~4.  As can be seen from eq.~(\ref{llw}), with $\Delta S_0=15$~J/mol/K fixed as before,  it is only $\rho_g$ now to be a free global parameter which now effectively also includes higher order effects in $\chi$ (specific copolymerization) and $\alpha$ (charging).  The data and fitting results are presented  in Fig.~4 with the best-fit $\rho_g$ shown in Tab.~\ref{Table_bestfit}.  We observe a very good performance of the  limiting law with an effective globule density $\rho_g=1.24$~mol/l. If there's physical reasoning  behind the good performance at even large $\alpha$ or low $c_s$, like cancellation of errors in our assumptions,  or if the good performance is just accidental, is difficult to judge.  However, at least for the  underlying data we find empirically that the simple form  eq.~(\ref{llw}) describes the whole dataset very well. 
Its usefulness for a more  general spectrum of data has to be  tested in future work. 

\begin{figure}[h!]
\vspace{0.3cm}
\begin{center}
\includegraphics[width=8.0cm,angle=0]{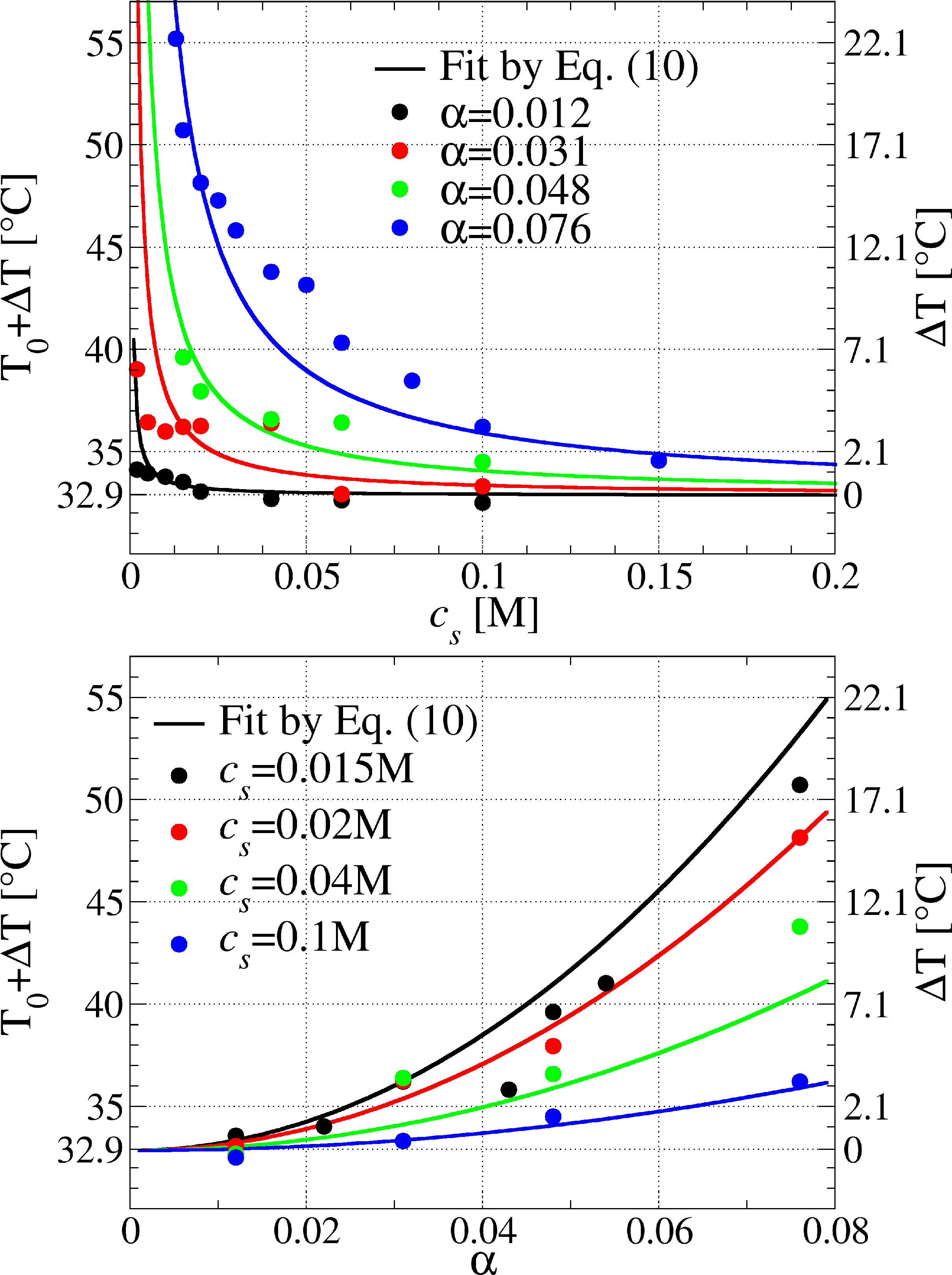}
\caption{LCST data based on the 90\% transmission criterion as in Fig.~3 but now the specific ($m$-value) ion effects are subtracted and full lines represent best theoretical fit based on the limiting law~(\ref{llw}).}
\label{T_09_fit_limit_figure}
\end{center}
\end{figure}

In summary, we have provided an analytical description for the changes of the LCST of charged copolymers
validated by novel, well-controlled cloud-point experiments of charged PNIPAM-based copolymers. The equations may serve as a guide for the development and optimization  of smart materials with desired properties, where the exact location of the LCST plays a decisive role. 
The general performance of our theoretical framework to other systems remains to be tested. 
We would like to note that we also aimed to compare to more available experimental data for various PNIPAM-based copolymers~\cite{Karbarz2006,Szczubialka,Kawasaki,KawasakiJPCB1997A,Spafford1998}. While all trends with $\alpha$ and $c_s$ could be  
qualitatively reproduced by our model, the  exact fitting was difficult due to the lack of detailed experimental key information, such as the exact salt
concentration, nature of the buffer, or pH or pK$_{\rm a}$ values of the monomers. Finally we note that our description may serve also for
a better interpretation of charge and salt effects on polypeptide systems, such as elastin.~\cite{cremer:elastin}

\section{Experimental Methods}

In our experiments the copolymers were prepared via
free ra\-di\-cal po\-ly\-me\-ri\-zation of NIPAM and an ionic liquid
mo\-no\-mer, 1-ethyl-3-vinyl\-imi\-da\-zo\-lium bro\-mide (EVImBr) as detailed in previous work.~\cite{Jiayin}
The solution stability of the copolymer was investigated through 
temperature-dependent UV/vis turbidity measurements at different 
salt concentrations $c_s$ between 0.01 and 1~M and four copolymer fractions $\chi=1.2\%$, 3.1\%, 4.8\%, and 7.6\%.  The raw transmission  data is presented in the SI. The LCST is then estimated by the 
location of the 'cloud point', that is, the temperature above which the polymer 
coils aggregate and precipitate and the dispersion becomes turbid as estimated
by light transmission loss. Furthermore, the pH was fixed to 7.0 in all measurements. As we are treating ionic liquid-like monomers without titratable hydrogens (no pK$_{\rm a}$ values), we can safely assume the fully charged state of this group,  that is, a valency of $z=+1$.  The absolute 
charge fraction is defined as $\alpha=|z\chi|$.  Changes of the LCST with pH are not a subject of this study.

\section{Acknowledgements}
J.D. thanks Felix Plamper for inspiring discussions and the Deutsche Forschungsgemeinschaft 
(DFG) for financial support.  J.H. and J.D. acknowledge funding from the Alexander-von-Humboldt 
(AvH) Stiftung, Germany. 


\providecommand*\mcitethebibliography{\thebibliography}
\csname @ifundefined\endcsname{endmcitethebibliography}
  {\let\endmcitethebibliography\endthebibliography}{}

\clearpage
\newpage

\begin{figure}[h]
\begin{center}
\includegraphics[width=8.0cm,angle=0]{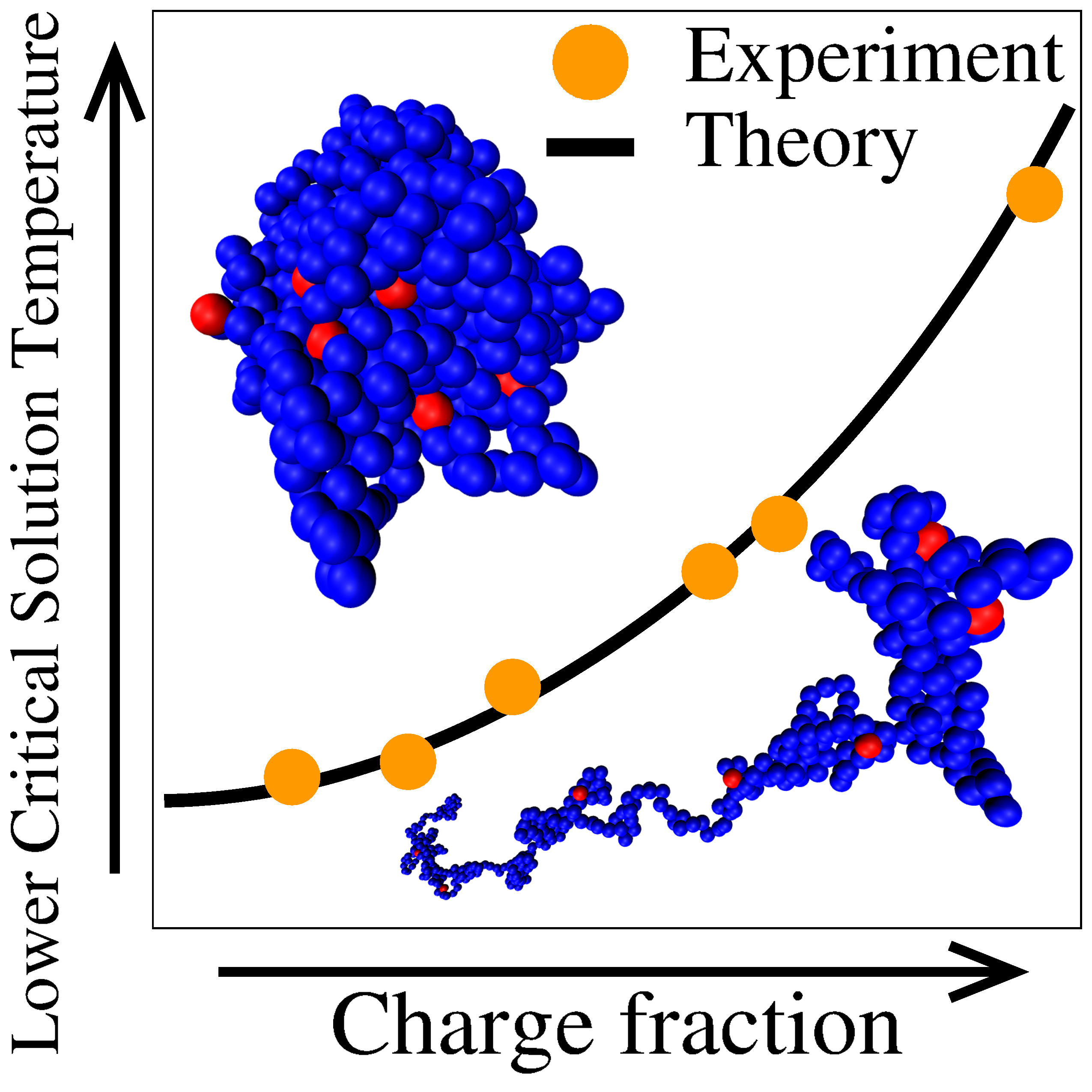}
\caption{TOC figure}
\label{TOC}
\end{center}
\end{figure}

\end{document}